# All-photonic entanglement swapping with remote quantum dots

Mattia Beccaceci[*,1], Giuseppe Ronco [*,†,1], Fabrizio Cienzo[1], Pierpaolo Bassetti[1], Alessandro Laneve[1], Francesco Basso Basset[2], Tobias M. Krieger[3], Qurin Buchinger[4], Francesco Salusti[5], Barbara Souza Damasceno[4], Silke Kuhn[4], Saimon F. Covre da Silva[6], Sandra Stroj[7], Klaus D. Jöns[5], Sven Höfling[4], Tobias Huber-Loyola[4,8], Armando Rastelli[3], Michele B. Rota[1], and Rinaldo Trotta [‡,1]

[1]Dipartimento di Fisica, Sapienza Università di Roma, Piazzale Aldo Moro 5, 00185 Roma, Italy
[2]Dipartimento di Fisica, Politecnico di Milano, Piazza Leonardo Da Vinci 32, 20133 Milano, Italy
[3]Institute of Semiconductor and Solid State Physics, Johannes Kepler University Linz, Altenberger Straße 69, 4040 Linz, Austria
[4]Technische Physik, University of Würzburg, Am Hubland, D-97074 Würzburg, Germany
[5]Institute for Photonic Quantum Systems (PhoQS), Center for Optoelectronics and Photonics Paderborn (CeOPP) and Department of Physics, Paderborn University, Warburger Straße 100, 33098, Paderborn, Germany
[6]Universidade Estadual de Campinas, Instituto de Física Gleb Wataghin, 13083-859 Campinas, Brazil
[7]Research Centre for Microtechnology FHV - Vorarlberg University of Applied Sciences Hochschulstraße 1, 6850 Dornbirn, Austria
[8]Institute of Photonics and Quantum Electronics (IPQ) and Center for Integrated Quantum Science and Technology (IQST), Karlsruhe Institute of Technology, Engesserstr. 5, 76131 Karlsruhe, Germany

**Abstract**

Entanglement swapping is a protocol that details how to create entanglement between previously uncorrelated particles. Its all-photonic version - mediated by the interference of photon pairs generated by separate quantum systems - finds disparate applications in quantum networks. So far, all-photonic entanglement swapping between remote systems has been implemented only using sources that operate probabilistically. However, the scaling up of quantum networks requires deterministic quantum emitters that do not suffer from a trade-off between degree of entanglement and photon-pair generation rate. Here, we demonstrate all-photonic entanglement swapping using photon-pairs generated by two separate GaAs quantum dots. The emitters are deterministically embedded in hybrid semiconductor-piezoelectric devices that make the entangled-photons from two dissimilar quantum dots nearly identical. Entanglement swapping is demonstrated with a fidelity as high as 0.71(2), more than 10 standard deviations above the classical limit. The experimental data are quantitatively explained by a theoretical model that also suggests how to boost the protocol performances. Our work opens the path to the exploitation of quantum dot entangled-photon sources in quantum repeater networks.

To develop a global quantum network, many challenges need to be addressed. One key undertaking is to efficiently generate and distribute entanglement across distant parties, bypassing the limits set by the no-cloning theorem[1]. This task requires quantum repeaters, i.e., "devices" that mitigate the exponential loss of quantum signals over distance by teleporting entanglement between previously uncorrelated nodes. There are a plethora of approaches to quantum repeaters[2,3], ranging from those based on matter quantum memories[4] to those relying on multi-photon graph states[5]. In many of them, swapping operations are used to share entanglement between remote network nodes by exploiting entanglement which is first generated locally[6–8]. Over the last decades, entanglement swapping has been implemented using different physical platforms[9–15], generating locally either atom(spin)-photon or photon-photon entanglement. The all-photonic version of entanglement swapping is particularly relevant for technological applications, as communication over long distances is made efficient by directly entangling flying qubits and by the straightforward possibility of concatenating swapping operations[3]. Moreover, efficient photon-photon entanglement swapping is also a crucial ingredient for photonic quantum computing[16]. Following the landmark experiment of Pan *et al.*[9], all-photonic entanglement swapping has been achieved in controlled laboratory settings[17], on a photonic chip[18], in the telecom band[19], and in deployed field infrastructure[20–22]. Yet, all these demonstrations rely on probabilistic photon sources, in which the probability of generating a single photon-pair per pulse can not be driven over the threshold imposed by the deleterious effect of multi-photon emission on entanglement. This fundamental constraint hampers their use in information processing, as the successful generation of entanglement between separate nodes may require a very large number of attempts, even after multiplexing. For this reason, implementations based on probabilistic photon sources can hardly be scaled up to the framework of practical quantum networks.

[*]These authors contributed equally
[†]giuseppe.ronco@uniroma1.it
[‡]rinaldo.trotta@uniroma1.it

Quantum emitters - such as atoms[23], quantum dots (QDs)[24], NV centers[25] as well as defects in 2D semiconductors[26] - address this hurdle by delivering on-demand single and entangled photons[27] with negligible multi-photon emission probability[28]. Quantum emitters can also host spin qubits, which can be entangled with the flying photons[29,30], can be used as quantum memories[13,31], and can enable the generation of complex multi-photon entangled states[32,33]. Despite pioneering swapping protocols have been implemented using spin-photon entanglement[10], or photon pairs produced by the very same quantum dot[34,35], to the best of our knowledge no experiment has demonstrated all-photonic entanglement swapping with separate quantum emitters. In this work, we implement this pioneering experiment and discuss its prospect for the construction of quantum networks.

## A quantum network based on quantum emitters

Figure 1a shows how to implement entanglement swapping with two ideal quantum emitters generating pairs of polarization entangled photons. The emission process occurs during the radiative cascade from an excited level $|E_2\rangle$ with angular momentum 0 to the ground state $|G\rangle$ through an intermediate, doubly degenerate level $|E_1\rangle$ with angular momentum ±1. This scheme produces maximally entangled $|\Phi^+\rangle$ Bell states and can describe both real[23] and artificial atoms[24]. It also embraces the general situation where the emitted photons feature different energies. The swapping protocol requires a Bell-state measurement (BSM) on two photons from each source - namely 2,3 in Fig. 1a -, which projects the state of the other photons, i.e. 1,4 in a maximally entangled state. This can be readily understood by rewriting the joint four-photon state in the Bell state basis:

$$|\Psi\rangle_{1234} = |\Phi^+\rangle_{12} \otimes |\Phi^+\rangle_{34} =$$
$$= \frac{1}{2}\Big(|\Phi^+\rangle_{14}|\Phi^+\rangle_{23} + |\Phi^-\rangle_{14}|\Phi^-\rangle_{23} +$$
$$+ |\Psi^+\rangle_{14}|\Psi^+\rangle_{23} + |\Psi^-\rangle_{14}|\Psi^-\rangle_{23}\Big). \quad (1)$$

From Eq. 1 it follows that a BSM on photons 2,3, projects the state of photons 1,4 in one of the Bell states, depending on the outcome of the BSM. The latter can be easily performed via two-photon interference using beam splitters and single photon detectors, albeit being able to detect only 2 out of 4 Bell states[36]. It is worth noting that the scheme of Fig. 1a is symmetric and entanglement can be swapped between photon 2,3 by performing the BSM on photons 1,4.

An entanglement-swapping chain can serve as a building block for quantum repeater networks and, as an example, we envisage here to use quantum emitters in protocols that rely on quantum memories[2]. Figure 1b shows a schematic of such a network, where multiple swapping operations are used to teleport entanglement between different nodes, each of them equipped with quantum memories. It is worth highlighting that the proposed scheme accounts for the different energies of the photons generated during the radiative cascades and, in addition, foresees the same energy for the photons traveling among the different nodes. This is relevant to establish a frequency-standard across remote network locations[37] and, in principle, allows the same type of quantum memory to be used for all the nodes. The implementation of such a network with quantum emitters presents a series of outstanding challenges, the first being the demonstration of entanglement swapping between remote sources.

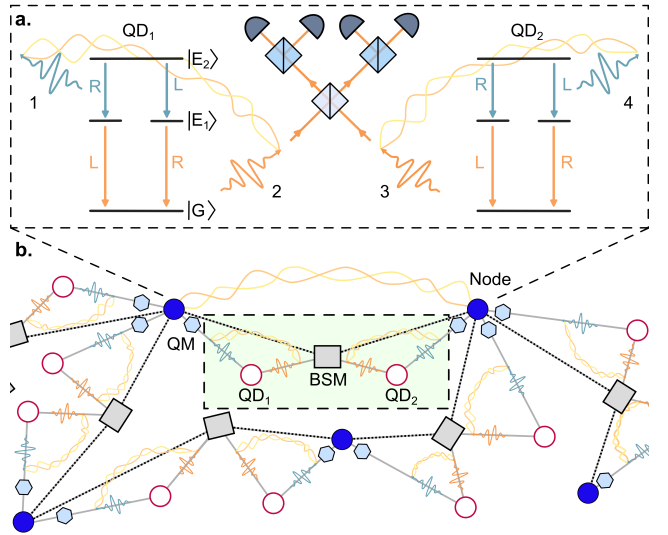

Figure 1: **Network-ready all-photonic entanglement swapping. a**, Quantum Dots (QDs) represented as two distinct two-level systems, acting as entangled photon sources. Each system emits two polarization-entangled photons via a radiative cascade initiated by the decay of the excited level $|E_2\rangle$ to the ground state $|G\rangle$, through an intermediate level $|E_1\rangle$. The two QD sources are used to implement entanglement swapping between photons 1,4 by performing a Bell state measurement (BSM) on photons 1,3, the latter requiring one beam splitter, two polarizing beam splitters, and single photon detectors. **b**, Conceptual quantum-network layout in which distinct entangled-photons sources and BSM stations are used to distribute entanglement across distant nodes. Each network node is equipped with quantum memories (QMs). We highlighted quantum (classical) channels with solid (dashed) line, while the dashed light-green rectangle marks the network block pictured in **a**.

Here, we exploit the biexciton (XX) – exciton (X) radiative cascade in QDs. This has been shown to generate deterministically highly-entangled photon pairs[38], with high indistinghuishability[39], record-low multi-photon emission probability[28] and - when integrated in photonic cavities[40] - high efficiency. Moreover, they can be interfaced with quantum memories[41], making them particularly suitable for the scheme of Fig. 1b. Despite these advances, there is one main challenge that has so far hampered the demonstration of all-photonic entanglement swapping: each QD generates photons with different energy, polarization, and temporal/spectral profile - a direct consequence of the spread in size, shape, and composition of different nanostructures. This feature prevents the use of distinct as-grown QDs to implement BSM based on the interference of indistinguishable photons. In fact, even the most sophisticated growth technique fails at controlling the QD structural details with atomic precision, being a single QD composed by several thousands of atoms. Post-growth tuning techniques have to be put in place to precisely reshape the emitted photons and, in the following, we detail the engineering steps that made the demonstration of all-photonic entanglement swapping with remote quantum dots possible.

## Engineering QDs for entanglement swapping

The device here exploited consists of a GaAs QD[38] obtained via droplet-etching epitaxy[42], and deterministically positioned at the center of a circular Bragg resonator (CBR)[40], in turn integrated onto a micro-machined $[Pb(Mg_{1/3}Nb_{2/3})O_3]_{0.72}$–$[PbTiO_3]_{0.28}$ (PMN-PT) piezoelectric actuator[27], see Fig. 2a. The underlying principle of this device is to use the strain fields generated by the PMN-PT to gain control over the energy of the entangled photons, while the CBR cavity reshapes their temporal/spectral profile.

The need to control the photon energy can be immediately recognized in Figs. 2b,c, which show photoluminescence (PL) spectra of two separate and dissimilar QDs (hereafter referred to as QD1 and QD2) as a function of the electric field $F_p$ applied to the PMN-PT. Without strain ($F_p = 0\,\text{kV/cm}$), the energy of both X and XX are different. More specifically, we find that the energy of the X (XX) photons generated by QD1 and QD2 differ by about $120\,\mu\text{eV}$ ($160\,\mu\text{eV}$), corresponding to 2.4 (2.2) times the linewidth extracted from Michelson interferometry (see Supporting Information, Sec. 2.3). This substantial detuning would hamper successful BSM and, as a consequence, entanglement swapping.

To recover for this detuning, we switch on the electric field $F_p$ across the PMN-PT below QD2 to bring the two X (or XX) transition energies in resonance with sub-$\mu$eV accuracy[43]. This is shown in Fig 2b,c, which also highlights

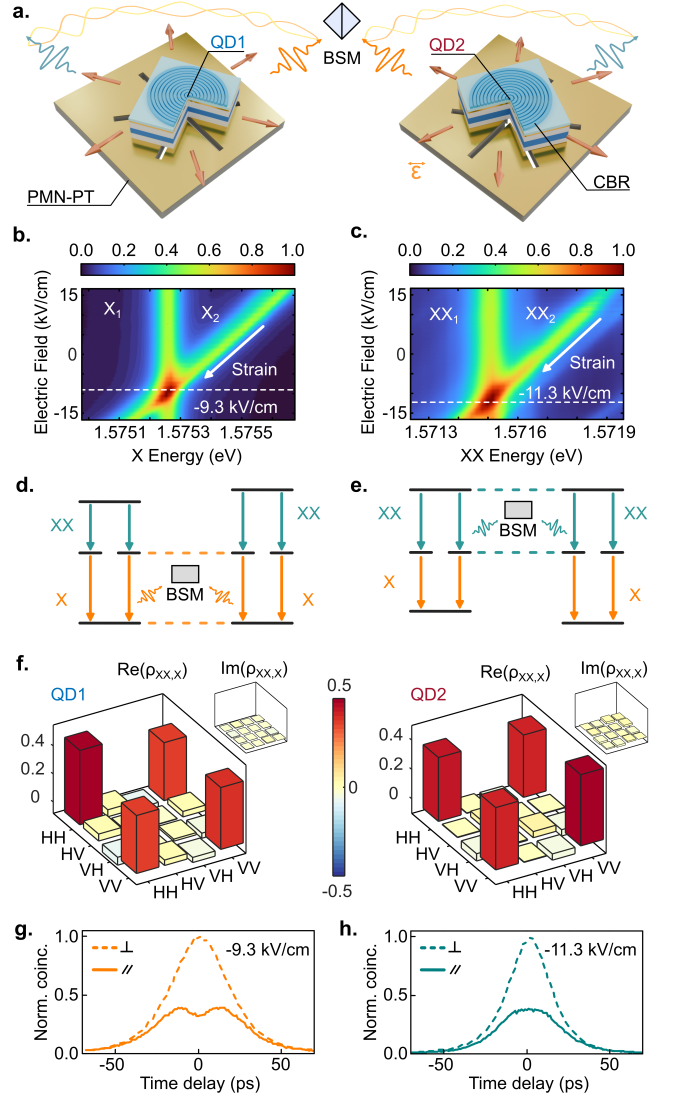

Figure 2: **Reshaping optical properties of remote QDs for entanglement swapping a**, Schematics of the interference between photons from two QDs embedded into a circular Bragg resonator (CBR) and integrated onto a piezoelectric actuator (PMN-PT). **b,c**, Colour-coded $\mu$-PL versus electric field $F_p$ showing energy tuning and crossings for X and XX. While QD1 is kept fixed, QD2 is tuned into resonance, achieved at $F_{p,X} = -9.3\,\text{kV/cm}$ and $F_{p,XX} = -11.3\,\text{kV/cm}$ for the X and XX photons, respectively. **d,e**, Remote entanglement swapping schemes. By means of strain tuning, it is possible to bring either the X (a) or the XX (b) of the two QDs in energetic resonance. In **c**, for clarity, we have artificially shifted the energy axis in one of the QDs so as to align the XX levels. **f**, Reconstructed density matrices of the XX-X photon-pairs for QD1 and QD2, respectively, at $F_{p,X}$. **g,h**, HOM coincidences histogram at the BS outputs for co- ($\parallel$) and cross-polarized ($\perp$) inputs around zero time delay and for both $F_{p,X}$ (panel **g**) and $F_{p,XX}$ (panel **h**)

two additional features. First, we notice that the overall tuning range is as large as 0.5 meV, meaning that it is possible to bring several QDs to emit at the same energy - not only two. Considering that the average spread in energy of the X transitions is about 3 meV, the probability of finding two QDs that can be strain-tuned to emit at the same energy is about 7 % (see Supporting Information, Sec. 4.1 for further details). Second, the Figs. 2b,c clearly show that when the X of the two QDs are brought into resonance ($F_{p,X} = -9.3$ kV/cm), the XX transitions still shows different energies. The same holds when the XX are tuned in resonance ($F_{p,XX} = -11.3$ kV/cm), i.e., the two X are found to be at different energy. This evidence is related to the dissimilar XX binding energies[44] of the two QDs, a consequence of the different Coulomb interactions among carriers therein confined.

The possibility of using $F_p$ to bring either the X or the XX photons to the same energy turns the swapping scheme of Fig. 1 into the two different configurations shown in Figs. 2d,e. More specifically, we can tune $F_p$ so as to implement entanglement swapping on the XX (X) photons by using the X (XX) photons to perform the BSM, as depicted in Fig. 2d (Fig. 2e). This occurs for $F_{p,X} = -9.3$ kV/cm and $F_{p,XX} = -11.3$ kV/cm, respectively. Once the QD energy levels are strain-tuned in one of these configurations, the quality of swapping critically depends on (*i*) the initial degree of entanglement of the photon pairs, and on (*ii*) the visibility of two photon interference.

(*i*) We assess the degree of entanglement of the X-XX photons generated by each QD by reconstructing the two-qubit density matrix via quantum-state tomography[45]. As quantitative measure of the degree of entanglement we make use of the fully entangled fraction $f = \max_\Phi \langle \Phi| \hat{\rho} |\Phi\rangle$, defined as the largest fidelity to a maximally entangled state[46] (for brevity we refer to it as entanglement fidelity). Figure 2f shows the reconstructed density matrix of the two QDs for $F_{p,X} = -9.3$ kV/cm, value for which the X energies are matched (see Figs. 2b,d). They closely resemble two $|\Phi^+\rangle$ states, with an entanglement fidelity of $f_1 = 0.90(2)$ and $f_2 = 0.91(1)$ for QD1 and QD2, respectively. In the Supporting Information, Sec. 4.3, we also show that the same degree of entanglement is measured when apply $F_{p,XX} = -11.3$ kV/cm to bring the XX energy in resonance (see Figs. 2e,f). In fact, for both QDs we find that the degree of entanglement remains practically constant with $F_p$ (see Supporting Information, Sec. 4.3), meaning that our device operate as a energy-tunable source of entangled photons[47]. This occurs even in the presence of a small residual splitting between the two X states thanks to the acceleration of the transition lifetimes induced by the cavity, see next point. Across the whole tuning range, the maximum value of the measured fidelity is $f$=0.93(1). This pinpoints to a small performance gap compared to the state of the art[38], likely related to the dynamical optical Stark shift[48] and the presence of a non-zero degree of polarization[49].

(*ii*) Having at hand the possibility to use strain to bring the photons generated by the two QDs in energetic resonance, the visibility of two-photon interference now critically depends on the temporal/spectral overlap of the photon wave-packets[50]. This, in turn, is connected to the interaction between each QD and the electromagnetic field confined in its own cavity - an effect that can be inferred via the Purcell factor[40]. It is therefore crucial to ensure that similar Purcell factors are obtained for different QDs, a task that requires extremely high accuracy in placing the QD in the center of the CBR cavity, which is also necessary to avoid undesired degree of polarization[51]. We tackle this challenge using a hyperspectral imaging technique[52] that enables spatial accuracy of about 15 nm and an overall device yield of about 70 %. Here, we highlight that the processing is performed after the transfer of the semiconductor membrane onto the piezoelectric actuator. Despite this additional challenge, we systematically observe high and similar Purcell factors in almost any QD (see Supporting Information, Sec. 3). For the X (XX) of QD1 and QD2, we measured similar Purcell of about 10 (8), with lifetimes as low as 25(5) ps (16(6) ps, see Supporting Information, Sec. 2.3 and Sec. 3).

To truly assess the indistinguishability of the photons generated by the two QDs we exploit Hong-Ou-Mandel interference. We perform the experiments when the XX or X energies are tuned by strain to implement the two swapping schemes of Figs. 2d,e. In both conditions, we collect coincidences between detectors placed at the output ports of a balanced BS, for co-($\parallel$) and cross-polarized ($\perp$) inputs (see Figs. 2g,h), prepared using a linear polarizer[43]. From these measurements, we extract an overall visibility $V$ of two photon interference of $V_X = 0.43(1)$ and $V_{XX} = 0.46(3)$, which increases to $V_X = 0.48(1)$ and $V_{XX} = 0.51(3)$ if we take into account the imperfections of the experimental setup and the non-zero value of the $g^{(2)}(0)$ (see Supporting Information, Sec. 4.3 for further details). Compared to previous results obtained for QDs in CBRs[53], this represents an improvement by about a factor 5. The visibilities are instead very similar to those obtained for QDs in bulk[54], thus proving that our nanophotonic cavity does not affect the optical quality of the QDs. Our visibilities are currently limited by the existence of energy-time entanglement between the photon pairs generated during the cascade[55] ad well as by the presence of residual charge noise[24]. Yet, owing to the sub-20 ps time resolution of our detection apparatus, we can improve the visibilities by temporally post-selecting events near zero-time delay[53] (see Supporting Information, Sec. 4.4). Using a time window of $\Delta t = 10$ ps, the X visibility - for which we can even resolve the dip at zero time delay due to the longer lifetime - increases up to $V_X = 0.66(1)$, while for the XX transition it reaches $V_{XX} = 0.61(1)$, without taking into account the setup imperfections and multiphoton emission contributions.

**Remote entanglement swapping**

Under the conditions described in the previous section, we perform the entanglement swapping protocols of Figs. 2b,c. We implement a BSM on the X (XX) photons to swap the entanglement to the XX (X) photons, on which we perform quantum state tomography. Figure 3a shows a schematic of the apparatus in the case of entanglement swapping applied to the XX photons (the protocol works identically for X photons). Considering Eq. 1, the outcome of the BSM on the X photons directly determines the XX Bell state. Specifically, if the BSM returns $|\Psi^-\rangle$ on the X, the XX are found to be in $|\Psi^-\rangle$ as well. As we implement the BSM using a BS followed by two polarization-selective BS (PBS, as shown in Fig. 3a), we are able to discriminate only $|\Psi^-\rangle$ (simultaneous detection at AB or CD) and $|\Psi^+\rangle$ (simultaneous detection at AC or BD) states[56]. This will return $|\Psi^-\rangle$ and $|\Psi^+\rangle$ states for the XX photons, respectively, which can be also evaluated as a function of the temporal window $\Delta t$ we use to post-select the BSM events.

Examples of the recorded density matrices for $\hat{\rho}^{\Psi^\pm}_{XX_1,XX_2}$ are shown in Figs. 3b,c for $\Delta t = 20\,\text{ps}$. They closely resemble $|\Psi^\pm\rangle$ states, clearly indicating successful entanglement swapping. Figure 3d also confirms that, when the tomography of the XX photons is performed without heralding on the outcome of the BSM, the resulting density matrix resembles a mixture of the four Bell states. We evaluate the entanglement fidelity of the XX photons as a function of $\Delta t$ and we display the obtained results in Fig. 3e. Remarkably, we find that the protocol is successful even without temporal post-selection, with a swapping fidelity as large as $f^{\Psi^-}_{swap,XX} = 0.60(1)$ ($f^{\Psi^+}_{swap,XX} = 0.56(1)$). This value exceeds the classical limit of 0.5[57] by 11 $\sigma$ (6 $\sigma$), unambiguously showing the first successful demonstration of all-photonic entanglement swapping with remote QD entangled photon-pair sources. Using temporal post-selection, the level of entanglement further increases up to $\overline{f}^{\Psi^-}_{swap,XX} = 0.71(2)$ and $\overline{f}^{\Psi^+}_{swap,XX} = 0.65(2)$), always well above the classical limit. As additional confirmation, no entanglement is measured when the QD spectra are detuned from resonance (see Supporting Information, Sec. 7.2).

We further demonstrate the protocol by performing the BSM on the XX photons and teleporting entanglement on the X ones, see Fig. 3e. Without temporal post-selection, we record $f^{\Psi^-}_{swap,X} = 0.62(2)$ ($f^{\Psi^+}_{swap,X} = 0.63(2)$), values that are slightly higher than those recorded for the XX photons due to the higher visibility of two photon interference, see Supporting Information, Sec. 4.4. Conversely, for $\Delta t = 20\,\text{ps}$ we obtain similar values within the error bars, i.e., $\overline{f}^{\Psi^-}_{swap,X} = 0.71(3)$ ($\overline{f}^{\Psi^+}_{swap,X} = 0.71(3)$).

In order to explain the experimental findings we develop a theoretical model for remote sources that predicts the swapping density matrices with no fitting parameters, but using the measured values of the lifetime, FSS, and HOM visibilities. The model also takes into account the specific protocol implementation - the use of the polarization-selective BSM as well as the temporal post-selection technique - and allows deriving an analytic form of the swapping matrix (see Supporting Information, Sec. 6). From the simulated density matrix, we compute $f_{swap}(\Delta t)$ without free parameters. The predictions, pictured as solid lines in Figs. 3c,d, quantitatively reproduce the experimental data and are even able to describe the superior performance for XX-based BSM at large $\Delta t$. The dashed line in Figs. 3e,f is the expected entanglement swapping fidelity taking into account the BS non-ideality and the multi-photon emission probability. After these corrections, we obtain a maximum $f_{swap}$ of 0.73(2) and 0.75(2), respectively for X and XX BSM.

Owing to the use of CBR cavities, the four-fold coincidence rates that we obtain are more than three orders of magnitudes larger than previous attempts focusing on photons from the very same QDs[34,35]. For both $|\Psi^\pm\rangle$ we measured swapping rates of a few Hz - see the top axis of Figs. 3e,f - which corresponds to a probability of establishing entanglement between uncorrelated photons equal to about $P_{swap} \sim 2 \cdot 10^{-5}$ (see Supporting Information, Sec. 5 for further details). Even after temporal post-selection, the four-fold coincidence rate remains close to 1 Hz, a direct consequence of the short lifetime of the X and XX transitions.

**Discussion and outlook**

Our first demonstration of all-photonic entanglement swapping between two independent and remote quantum emitters marks a major step towards the construction of quantum repeater networks with deterministic entangled photon-pairs sources. Within the broader vision of realizing the quantum network displayed in Fig. 1b, future developments should pursue three key objectives: i) Boosting both the entanglement swapping fidelities and rates; ii) Interfacing the swapped photons with suitable quantum memories capable to store and retrieve polarization qubits; iii) Deploying the developed technology in urban quantum communication scenarios.

(*i*) Even if the swapping fidelities are unambiguously above the classical limit, advanced quantum communication applications requires higher level of entanglement. As benchmark, we take a yet-to-be demonstrated quantum key distribution (QKD) protocol with entanglement-swapped photons from remote QDs. To successfully share a secret key between remote network locations, we estimate that it is necessary to have an entanglement fidelity well above 80 %. This is higher than what we achieved here, but our theoretical and experimental work suggests the path for improvement:

considering the high entanglement of the initial photon pairs, the swapping fidelity is primarily limited by the accuracy of the BSM and, therefore, by the visibility of two-photon interference, as mentioned above.

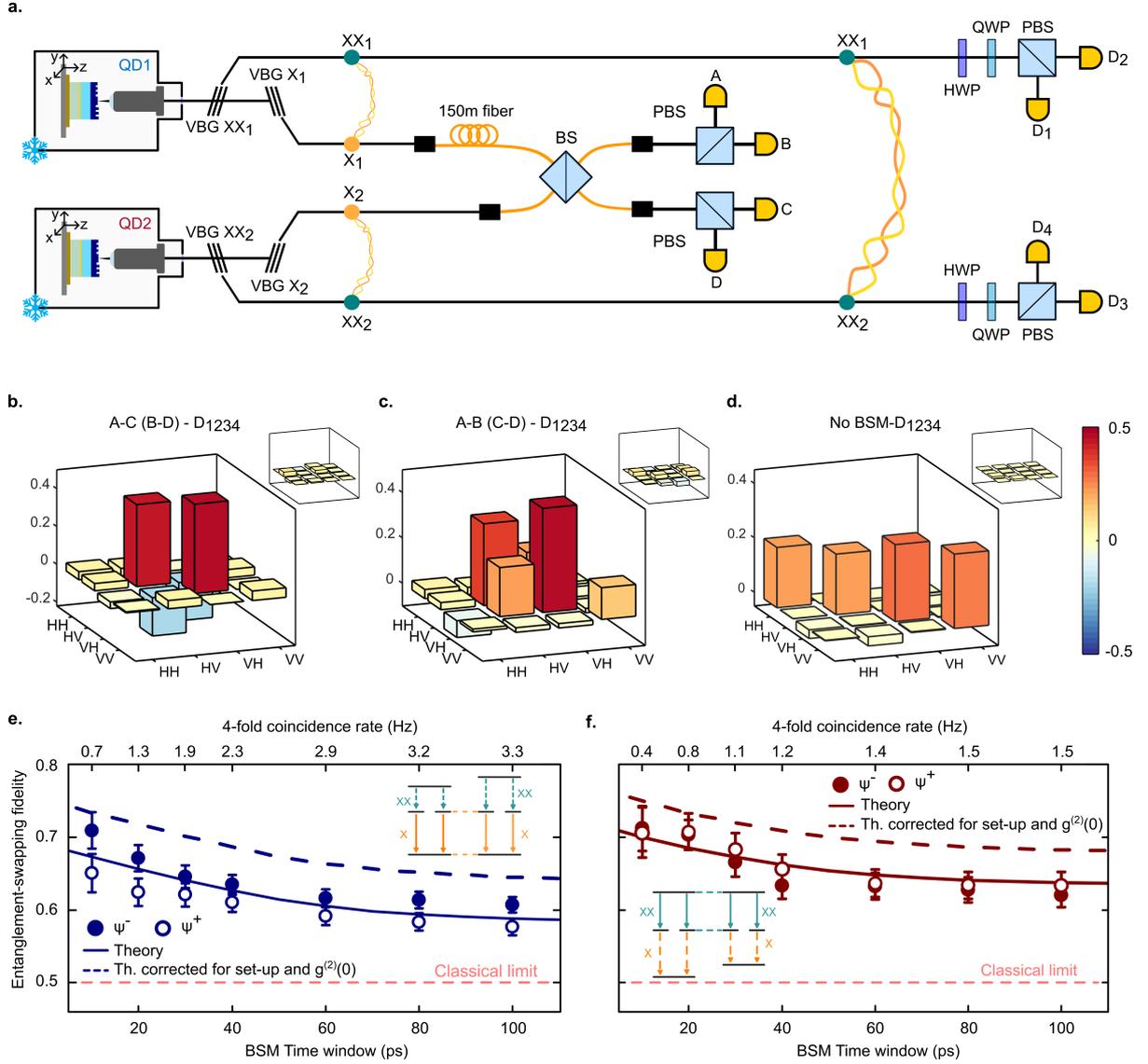

Fig. 3: **Entanglement swapping between remote QDs. a**, Schematic of the swapping setup. The two sources are kept in separate cryostats, and the X and XX emission is collected, redirected, and spectrally filtered trough a series of mirrors and Volume-Bragg Gratings (VBGs). Then, the X photons are directed into a BSM setup, and the XX photons are guided to an apparatus for quantum state tomography. The same setup can be used to perfom the BSM on the XX photons so as to swap entanglement between the X photons. Different combinations of gated coincidences of detectors $D_{1,2,3,4}$ with A,B,C,D provide density matrix for both $|\Psi^+\rangle$ and $|\Psi^-\rangle$ states. **b, c**, Real and imaginary part (inset) of the reconstructed density matrices for the swapped $XX_1$–$XX_2$ photons, conditioned on $|\Psi^-\rangle$ and $|\Psi^+\rangle$ BSM events, respectively. **d**, Real and imaginary part (inset) of the same density matrix as obtained without gating on the outcome of the BSM, i.e., on detectors A,B,C,D. The matrix is diagonal, as expected from an incoherent superposition of Bell states, and shows no degree of entanglement. **e**, Entanglement swapping fidelity $f_{swp}$ of $|\Psi^\pm\rangle$ versus the BSM time window $\Delta t$. The solid line shows the expected values from our theoretical simulations. The dashed line describes the values that could be ideally obtained after correcting for the $g^{(2)}(0)$ and the setup imperfections (see Supporting Information, Sec. 6 for further details). The classical limit of 0.5 - red dashed line - is well surpassed even without temporal post-selection. The top axis shows the four-fold coincidence rates. The inset shows the energy level scheme used to implement the swapping experiment. **f**, Same as **e**, but for an experiment in which entanglement is swapped between the X photons by performing the BSM on the XX photons.

Our visibilities are currently limited by the existence of energy-time entanglement[55] as well as by the presence of residual charge noise[24].

To alleviate for this hurdle, one can engineer the ratio between the X and XX lifetime either by designing photonic cavities with differential Purcell enhancement for X and XX[55], or by using diode-like structures to control the electric field across the QD[58]. The latter strategy has demonstrated HOM visibilities as high as 80 % without temporal postselection, a value that would allow pushing the swapping fidelity just above the threshold needed to successfully perform QKD. The electric field can be also used to suppress residual charge noise, with demonstrated HOM visibilities as high as 92 %[59], albeit in structures with negligible Purcell enhancement and without pumping the XX state for entangled photon generation. The electric field can be also instrumental to increase the overall swapping rates. As discussed in the Supporting Information, Sec. 5, the presence of blinking in our CBR limits the fourfold coincidence rate by almost a factor 80. However, recent results[58] have demonstrated that blinking can be fully suppressed using electric fields. Therefore, we envisage that the use of diode-like structures in properly designed CBRs will be instrumental to boost the level of entanglement and minimize the number of attempts needed to successfully establish entanglement between remote network nodes. More specifically, in the Supporting Information, Sec. 5, we estimate that state of the art improvements of the source can push $P_{swap}$ up to 0.17. This value, when using our experimental set-up, would result in a coincidence rate close to 3 kHz, already above the value that can be achieved with probabilistic sources of entangled photons[19].

(*ii*) Interfacing single and entangled photons from QDs with quantum memories is crucial for a functioning quantum network. There are several systems that can be used for this purpose, such as rare-earth–doped, diamond color centers, and alkali-based quantum memories[41]. The latter have been used in recent demonstrations of on-demand storage and retrieval of single photons from QDs[60], even in combination with CBR cavities[61]. Despite these advances, additional efforts are needed to improve the memory efficiency, the storage time, and extend its operation to polarization-entangled photons. Modifications of the source are also needed, as bandwidth matching is crucial to boost the overall protocol efficiency. The differential Purcell enhancement proposed above is particularly suited for this purpose, as the (short) XX photons could be used for the BSM while the (long) X photons could be interfaced with alkali atoms. This type of memory is also interesting for setting a frequency standard across the network (see Fig. 1b). In this case, it would be also necessary to erase the small differences in XX binding energies we observe in dissimilar QDs (see Figs. 2d,e and Supporting Information, Sec. 4.3), a task which can be accomplished using the combined effect of strain and electric field[44]. This would allow turning the swapping schemes of Figs. 2b,c into the universal one shown in Fig. 1a. Lastly, we want to point out that recent progress on spins confined in GaAs QDs[31] show that quantum memories based on exactly the same hardware shown here could become feasible, with one quantum dot acting as entangled-photons source, and another as a memory.

(*iii*) The next step is the exploitation of our entangled photon sources in real-life quantum networks. Given the emission wavelength of our GaAs QDs, free space quantum links are the ideal choice. Recent demonstrations of entanglement-based QKD[48,56] as well as quantum teleportation[53] suggests that remote entanglement swapping in urban communication scenario is within reach. Once the source is improved, satellite entanglement-based quantum communication[62] is also a path to explore. That said, we point out that our results are relevant also for fiber-based quantum communication. Recent works have shown enormous progress in the performance of single[63] and entangled photons sources[64] based on C-band QDs, and our results could be directly transferred to the telecommunication bands. Alternatively, it is also possible to use GaAs QDs in combination with frequency-converters, as recently used to implement quantum teleportation[65]. Therefore, we envisage that our first demonstration of entanglement swapping between remote QDs will now open the path for the construction of large scale quantum networks that exploit deterministic entanglement resources to distribute entanglement across distant parties.

## Methods

### Optical Characterization

The two QDs are hosted at cryogenic temperature (3.5 K and 5 K, respectively) in two independent closed-cycle cryostats. Excitation and collection are performed in back-reflection geometry with a 0.5-NA aspheric lens, used both to focus the pump and to collect the signal. A Ti:sapphire laser provides 140 fs excitation pulses for both sources at a repetition rate of 80 MHz, doubled using a Mach-Zender interferometer. A 4f pulse slicer selects a narrow spectral component, with bandwidth tunable from ∼ 80 $\mu$eV to a few hundred $\mu$eV. The bandwidth is chosen to be long enough in the time domain to mitigate the optical Stark shift[48], yet spectrally sharp to avoid laser photons overlapping the X and XX energies. Spectral filtering based on the reflection of Volume-Bragg Gratings (VBGs) isolates the X and XX transitions, which are subsequently directed either to the BSM stage or to the quantum-state-tomography setup through optical fibers. For FSS measurements, a half-wave plate and a polarizer are used; the half-wave plate is rotated by $\pi$ at steps of 2° while synchronizing the acquisition of a CCD camera placed at the output of a 750-mm spectrometer. In combination with an 1800-lines/mm grating, the setup achieves a spectral resolution of about 0.02 nm.

**Polarization-selective BSM**

Thanks to polarization-selective optics, the BSM setup enables discrimination between both $|\Psi^\pm\rangle$ Bell states. The apparatus, schematically illustrated in Fig. 3a, consists of a 50:50 fiber BS for HOM interference. The two BS outputs are collimated and directed into free space through two equivalent optical paths, each comprising a PBS followed by a quarter - half - quarter wave plate sequence for correcting polarization rotations. The four PBS outputs are fiber-coupled and sent to superconducting nanowire single-photon detectors (SNSPDs) featuring an average timing jitter of 15 ps and connected to ultra-fast electronics with 2 ps time resolution. A coincidence event between the two outputs of the same PBS corresponds to a $|\Psi^+\rangle$ detection, whereas coincidences between orthogonal outputs of different PBSs identify $|\Psi^-\rangle$ events[56]. These coincidences clicks serve as gate counts for gating the quantum state tomography on the remaining photons.

**Hong-Ou-Mandel interference**

HOM interference between either X or XX photons occurs at the fiber-BS of the BSM setup. The two photons are sent through single-mode fibers to interfere at the BS, and the outputs are directly connected to the SNSPDs. The excitation pulse from a common laser is divided into two paths using a fiber BS, with one arm including a delay line. By monitoring the coincidence counts between the two photons, we temporally matched the excitation pulses so that the photons arrived simultaneously at the BS. To obtain the histograms shown in Figs. 2h,i, we selected a specific component of the FSS and rotated its polarization to align with a common reference axis for both sources. In this configuration, interference occurs only when the photons are indistinguishable. The HOM visibility - defined as the relative depth of the coincidence dip at zero time delay for parallel-polarized photons with respect to the coincidences for orthogonally polarized photons - was extracted from coincidence histograms acquired under both polarization settings. Since the visibility is derived from the integrated coincidences around zero delay, we post-selected the events within a time window centered at zero-time delay $\tau = 0$ and integrated the coincidence counts over the interval $[-\Delta t, \Delta t]$. The visibility is therefore given by

$$V(\Delta t) = 1 - \frac{C_\parallel^{\Delta t}}{C_\perp^{\Delta t}}, \quad (2)$$

where $C_{\parallel,\perp}^{\Delta t}$ represent the integrated coincidence areas within the selected window. Note that the effective HOM visibility contributing to the BSM operation is lower than the directly measured value, due to chromatic dispersion arising from the finite FSS (see Supporting Information, Sec. 4.4).

**Acknowledgements**

The authors thank Paolo Mataloni for the fruitful discussions. This project has received funding from the European Union's Horizon 2020 research and innovation program under Grant Agreement no. 899814 (Qurope) and No. 871130 (Ascent+), and from the QuantERA II program that has received funding from the European Union's Horizon 2020 research and innovation program under Grant Agreement No 101017733 via the project QDE-QKD and the FFG grant no. 891366, and from the European Union's Horizon Europe research and innovation program under EPIQUE Project GA No. 101135288. The authors also acknowledge support from MUR (Ministero dell'Universita e della Ricerca) through the PNRR MUR project PE0000023-NQSTI, and the European Commission by project QUID (Quantum Italy Deployment) funded in the Digital Europe Programme under the grant agreement No 101091408. AR acknowledges support of the EU HE EIC Pathfinder challenges action under grant agreement No. 101115575 (Q-ONE), from the QuantERA program via the project MEEDGARD (FFG Grant No. 906046), the Austrian Science Fund (FWF) [10.55776/COE1 , FG5, PIN4389523] via the Research Group FG5 [10.55776/FG5], and the cluster of excellence Quantum Science Austria (quantA) [10.55776/COE1], as well as the Linz Institute of Technology (LIT), and the LIT Secure and Correct Systems Lab, supported by the State of Upper Austria. THL acknowledges Funding from the German Federal Ministry of Research, Technology and Space (BMFTR) through the project Qecs (FKZ: 13N16272). This work is supported by the Deutsche Forschungsgemeinschaft (German Research Foundation) through the transregional collaborative research center TRR142/3-2022 (231447078) and the European Research Council starting grant (LiNQs,101042672). S.F.C. da Silva acknowledges São Paulo Research Foundation (FAPESP), Brasil, Process Number 2024/08527-2 and 2024/21615-8 for financial support.

**Author contributions**

M. B. and G. R. performed the experiment with the contribution of F. C., P. B., and A.L. under the supervision of M. B. R. and R. T.. M. B. R. developed the data acquisition